# Polarization manipulation of giant photonic spin Hall effect using wave-guiding effect

Monu Nath Baitha and Kyoungsik Kim[*]

*School of Mechanical Engineering, Yonsei University, 50 Yonsei-ro, Seodaemun-gu, Seoul 03722, Republic of Korea.*

[*]Correspondence and requests for materials should be addressed to K.K. (email: kks@yonsei.ac.kr).

**ABSTRACT**

In plasmonic systems, the enhanced photonic spin Hall effect (PSHE) was previously possible only for horizontal polarization. By employing the wave-guiding surface plasmon resonance (WG-SPR) effect, we report a giant photonic spin Hall effect (G-PSHE) of reflected light for both horizontal and vertical polarization waves. We investigated the polarization-manipulated G-PSHE in the Kretschmann configuration with an additional glass dielectric layer. This additional dielectric layer allowed us to achieve millimeter-scale (more than 2 mm to sub-millimeter) G-PSHE. We achieved polarization manipulation by designing novel structures employing wave-guiding and SPR theory. Using a simulation study, we investigated the impact of an additional thin dielectric layer on G-PSHE. This study enables the potential application of both horizontal and vertical polarization-based quantum devices and sensors for which light spin plays a pivotal role.

## 1. Introduction

When a light beam interacts with an interface consisting of two different media, there are some deviations, spatial and angular Goos-Hänchen (GH) shifts, and spatial and angular Imbert-Fedorov (IF) shifts[1-8]. The spatial IF shift is the phenomenon of separation of the incident plane polarization wave to left-handed and right-handed circularly polarized components in opposite directions. This spatial IF shift is perpendicular to the incident plane and is also known as the photonic spin Hall effect (PSHE)[8-10]. This PSHE originates from the spin-orbit interaction (SOI) of light, the geometric Berry phase, and the conservation of angular momentum of light[9-11]. This SOI appears due to the coupling of internal (polarization) and spatial or orbital (phase distribution and intensity) degrees of freedom of light. The SOI plays an important role in describing the photonic Hall effect of light. Bliokh et al. provided a complete theoretical explanation of the SOI of light and also showed the PSHE of the paraxial beam after reflection/refraction from an interface[9-11]. Recently, PSHE has been a prime focus owing to its promising prospects and applications in spin optics, resulting in an exponential increase in the number of publications and patents in the past few years.

The first concept of PSHE was theoretically introduced by Onoda et al.[10] and was experimentally confirmed by Hosten and Kwiat for an air-glass interface[12]. PSHE has been explored for many different models and structures, such as nanostructures[13], metal films[14], magnetic films[15], graphene[16, 17], metamaterials[18], surface plasmonic resonance (SPR)[19, 20], near Dirac point[21], gradient refractive index media[22, 23], gratings[24, 25], liquid crystals[26], dielectric spheres[27], uniaxial birefringent crystals[28], metasurfaces[29], and hyperbolic metamaterials[30, 31]. PSHE paves the way for the development of potential applications in the real world, ranging from spin-dependent beam splitters[32] to surface sensors[17, 33].

As the PSHE usually appears on the scale of a few nanometers, it is difficult to measure in the lab setup by any direct method. The enhancement of PSHE phenomena is important to develop a range of light-spin-based potential application devices. Enhanced PSHE was observed near the Brewster angle[34, 35] and in multilayer nanostructures[31, 36, 37], metal cladding waveguide structures[38], and anisotropic impedance mismatching methods[39].

In this letter, we report a theoretical analysis of the giant photonic spin Hall effect (G-PSHE) through the wave-guided surface plasmon resonance (WG-SPR) effect. We consider a Kretschmann configuration with an additional glass dielectric layer attached to a ultrathin gold plasmonic layer. The enhancement in G-PSHE is on the order of $10^6$ nm (a few millimeters to sub-millimeter), which is significantly larger than that in the previously reported guided-wave surface plasmon resonance model[40]. Furthermore, as a unique accomplishment, we have shown the PSHE enhancement for both horizontal and vertical polarization by utilizing an extra dielectric layer, resulting in a wave-guiding effect, which differs from the previously published plasmonic assisted PSHE study. By employing this wave-guiding design, we can also control the active polarization mode and the enhancement of G-PSHE.

## 2. Theoretical Method

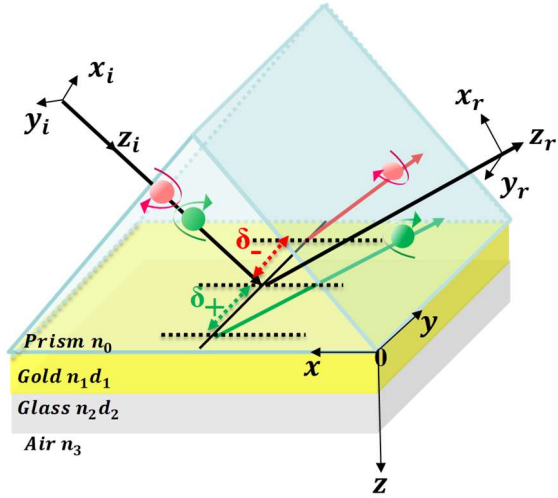

**Fig.1.** Schematic representation of spin Hall effect of reflected light from the prism-coupled wave-guiding surface plasmon resonance model, where $\delta_+$ and $\delta_-$ represent the transverse shift from the center positions of the left-hand and right-hand circularly polarized waves, respectively.

The most common way to generate SPR is the Kretschmann configuration, in which a nanometal film is deposited on the base of the prism. Transverse magnetic (TM) waves can only excite surface plasmon resonance (SPR), which generates an evanescent field near the metal surface. Therefore, a small change in the refractive index near the metal surface results in a substantial change in the SPR dips[41].

The WG-SPR model has two parameters for a double-layer structure coupled with a prism in the air. The first layer is made of gold (Au) with thickness $d_1$ at the bottom of the coupled prism, and the second layer is glass with thickness $d_2$ attached to the gold layer, as shown in Fig. 1. For simulation and calculation purposes, the refractive index of the prism ($n_0$) and glass ($n_2$) layer is taken as $n_0 = n_2 = 1.515$, and that of gold is taken as $n_1 = 0.2166 + 3.2322i$ at a wavelength of $\lambda_0$=632.8 nm. The refractive index $n_3$ of the air substrate was considered to be 1.

A plane-polarized incident wave is decomposed into left-handed and right-handed circular polarization with shifts opposite to each other and perpendicular to the incident plane. A monochromatic beam is considered to be incident through a prism on the surface of the WG-SPR structure in our study. The Cartesian coordinate system $(x, y, z)$ is fixed to WG-SPR, and the prism-Au interface plane is at $z=0$. The incident and reflected beams are attached to the laboratory Cartesian coordinate frame with coordinates $(x_i, y_i, z_i)$ and $(x_r, y_r, z_r)$, respectively. In the spin bias set representation, the incident monochromatic wave can be observed in two different polarization states.

$$\widetilde{E}_i^H = \frac{1}{\sqrt{2}}\left(\widetilde{E}_{i|+>} + \widetilde{E}_{i|->}\right)$$
$$\widetilde{E}_i^V = \frac{i}{\sqrt{2}}\left(-\widetilde{E}_{i|+>} + \widetilde{E}_{i|->}\right) \quad (1)$$

where H and V represent the horizontal and vertical polarized waves, respectively. The horizontal polarization is within the plane of incidence, and the vertical polarization is out of the plane. $\widetilde{E}_{i|+>}$ and $\widetilde{E}_{i|->}$ represent the left-handed and right-handed circularly polarized waves, respectively. The spectrum of the incident light is extremely narrow and is given as

$$\widetilde{E}_{i|\pm>} = (\vec{e}_{ix} + i\sigma\vec{e}_{iy})\frac{\omega_0}{\sqrt{2\pi}}e^{-\left(\frac{\omega_0^2(k_{ix}^2+k_{iy}^2)}{4}\right)} \quad (2)$$

where $\omega_0$ is the light waist and $\sigma = \pm 1$ denotes plane operators, where $\sigma = +1$ and $\sigma = -1$ are left circularly polarized light and right circularly polarized light, respectively. Owing to the matching of the boundary conditions, the reflected light can be represented as

$$\widetilde{E}_r^H = \frac{r_p}{\sqrt{2}}\left[e^{(ik_{ry}\delta_{r|+>}^H)}\widetilde{E}_{r|+>} + e^{(-ik_{ry}\delta_{r|->}^H)}\widetilde{E}_{r|->}\right] \quad (3)$$

$$\widetilde{E}_r^V = i\frac{r_s}{\sqrt{2}}\left[-e^{(ik_{ry}\delta_{r|+>}^V)}\widetilde{E}_{r|+>} + e^{(-ik_{ry}\delta_{r|->}^V)}\widetilde{E}_{r|->}\right] \quad (4)$$

The term $e^{(\pm ik_{ry}\delta_{r|\pm>}^{H,V})}$ is responsible for the spin-orbit interaction of light, where

$$\delta_{r|\pm>}^H = \mp\frac{\lambda_0}{2\pi}\left[1 + \frac{|r_s|}{|r_p|}\cos(\varphi_s - \varphi_p)\right]\cot\theta_i \quad (5)$$

$$\delta_{r|\pm>}^V = \mp\frac{\lambda_0}{2\pi}\left[1 + \frac{|r_p|}{|r_s|}\cos(\varphi_p - \varphi_s)\right]\cot\theta_i \quad (6)$$

In the above expression, $r_{s,p}$ are the Fresnel reflection coefficients for the |s> and |p> polarized incident waves, respectively, and $\varphi_{s,p}$ are the corresponding phases. $\delta_{r|\pm>}^H$ and $\delta_{r|\pm>}^V$ are measures of the shifts of the left-and right-handed circular polarization components of the polarized light of H and V, respectively. The derivation is adopted from previous publication[19] and a detailed discussion on the relevant formulas is provided in Appendix-A

Under the theoretical assumption, the wavelength of light ($\lambda_0$) is much smaller then the beam waist ($\omega_0$). The relationship of both, $\omega_0^2 \gg (\lambda \cot\theta_i/2\pi)^2$, is referred as large beam waist condition[42]. This large beam waist condition is not valid for strongly confined light or near normal incident wave[38]. In general the PSHE is much smaller than the beam waist but numbers of potential way have been found and presented to get PSHE with multiple fold of incident wavelength[38, 40, 43, 44]. The presented study shows a giant PSHE due to strong field confinement.

The physical mechanism behind the PSHE phenomenon is the SOI. Originating from spin-orbit coupling after interaction with the material, mutual influence between the light trajectory and polarization (spin) of the light beam occurs. In Eqs. (3) and (4), the term $e^{(\pm ik_{ry}\delta_{r|\pm>}^{H,V})}$ represents the spin-orbit coupling, and the ratio $|r_s|/|r_p|$ ($|r_p|/|r_s|$) plays a significant role in enhancing the spin–orbit interaction, which ultimately results in the enhancement of PSHE. Therefore, to gain G-PSHE, the denominator Fresnel coefficient for each case typically approaches zero.

This very small amount of PSHE is very difficult to realize in the laboratory; therefore, researchers are looking forward to enhancing and controlling this effect. Recently, Dai et al. experimentally showed ultrahigh-order modes of PSHE[38], and Wen and Zubairy showed tunable PSHE using an N-type coherent SPR medium[45]. To further enhance G-PSHE, we adopted the WG-SPR model in which a glass dielectric wave-guiding layer with an optimized thickness comparable to the operating wavelength is incorporated between the thin Au layer and air. We investigated the influence of this glass layer on the enhancement of G-PSHE with polarization manipulation.

## 3. Result and Discussion

By using Eqs. (5) and (6), the PSHE was calculated for the general Kretschmann configuration presented in Fig. 1. In our study, the thickness of the gold thin film ($d_1$) was considered under its skin depth to enable the easy transmission of incident waves into the wave-guiding layer from 1 nm to 30 nm[46]. Horizontal polarized ($\delta_+^H$) and vertically polarized ($\delta_+^V$) PSHEs with respect to the incident angle ($\theta_I$) were calculated without a wave-guiding layer for varying Au layer thicknesses, as shown in Fig. 2(a).

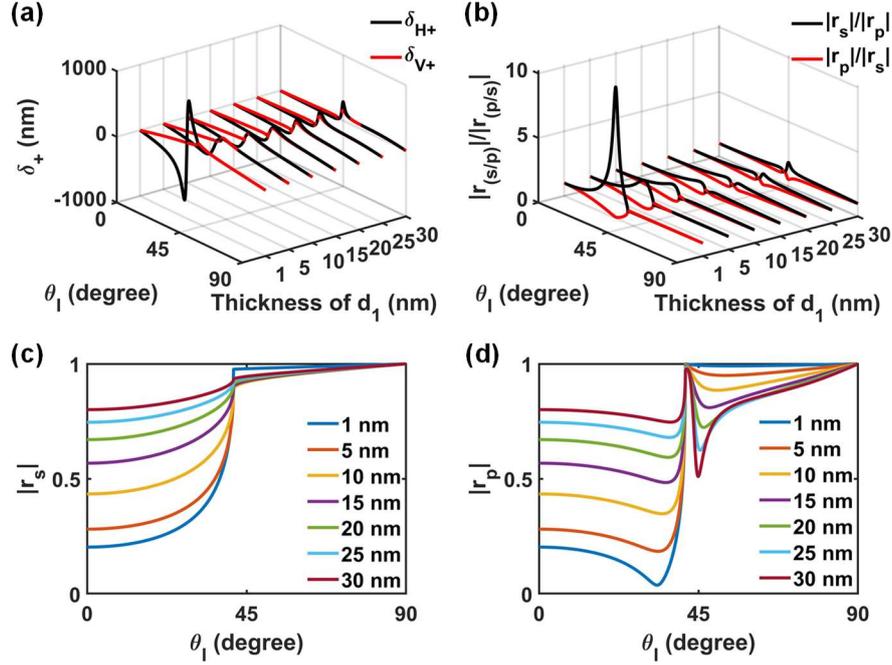

**Fig. 2.** (a) In the case with no additional glass layer, i.e., $d_2=0$, we calculated the photonic spin Hall effect of horizontal polarization $\delta_+^H$ and vertical polarization $\delta_+^V$, (b) absolute Fresnel reflection coefficient ratio $|r_s|/|r_p|$ and $|r_p|/|r_s|$, and absolute Fresnel reflection coefficients (c) $|r_s|$ and (d) $|r_p|$ for the incident $|s\rangle$ polarized wave, and $|p\rangle$ polarized wave versus the incidence angle for the different Au thickness ($d_1$ = 1 nm, 5 nm, 10 nm, 15 nm, 20 nm, 25 nm, and 30 nm).

Fig. 2 represents the calculation for standard Kretschmann configuration, when there is only metal layer without dielectric. Only p-polarized incident waves with a resonance angle are used in the popular Kretschmann configuration for local field enhancement, i.e. SPR. Because of this, there is a resonant dip for p-polarized waves only, resulting in a greater PSHE for H-polarized waves compared to V-polarized waves. In Fig. 2(a), the horizontal polarized PSHE ($\delta_+^H$) is larger than the vertically polarized PSHE ($\delta_+^V$), while both are reduced with an increase in the Au thickness ($d_1$). The PSHEs for the 1 nm-thick Au layer were $\delta_+^H$ ~800 nm and $\delta_+^V$ ~230 nm. For a 30 nm-thick Au layer, $\delta_+^H$ and $\delta_+^V$ are ~280 nm and ~150 nm, respectively. This result is consistent with the previously reported PSHE for thin Ag films[47].

As shown in Eqs. (5) and (6), one of the key factors of PSHE is the ratio of the absolute Fresnel reflection coefficient, that is, $|r_s|/|r_p|$ for horizontal polarization and $|r_p|/|r_s|$ for vertical polarization. When the Au layer is 1 nm, $|r_s|/|r_p|$ is a maximum of ~10 for horizontal polarization and $|r_p|/|r_s|$ is a maximum of ~1 for vertical polarization (Fig. 2(b)). As the Au thickness ($d_1$) increases, the ratios $|r_s|/|r_p|$ and $|r_p|/|r_s|$ decrease. The absolute Fresnel reflection coefficients $|r_s|$ and $|r_p|$ of the corresponding $|s\rangle$ and $|p\rangle$ polarized incident waves are calculated, as shown in Figs. 2(c) and 2(d).

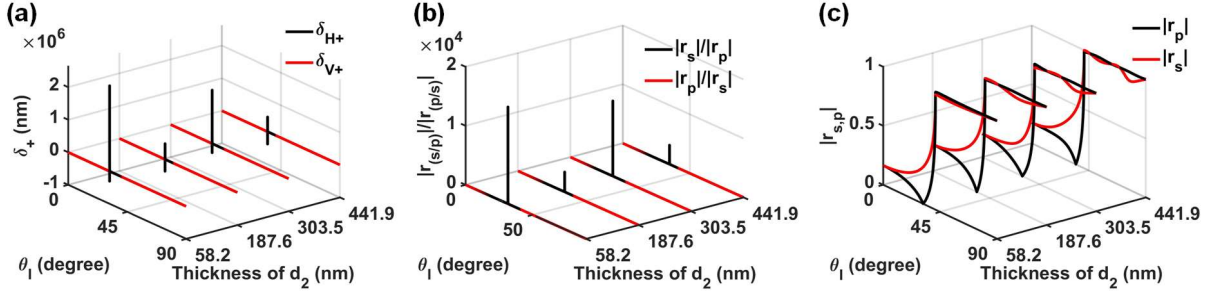

**Fig. 3.** (a) When the Au layer ($d_1$) is 1 nm thick, by changing the thickness ($d_2$) of a wave-guiding glass layer (58.2 nm, 187.6 nm, 303.5 nm, 441.9 nm), we calculated the photonic spin Hall effect of horizontal polarization $\delta_+^H$ and vertical polarization $\delta_+^V$ versus the incidence angle. (b) The absolute Fresnel reflection coefficient ratios $|r_s|/|r_p|$ and $|r_p|/|r_s|$. (c) The absolute Fresnel reflection coefficients $|r_s|$ and $|r_p|$ for the incident $|s\rangle$ polarized wave and $|p\rangle$ polarized wave with respect to incident angle.

Fig. 3 illustrates the proposed WG-SPR model; the Au thickness ($d_1$) is kept constant at 1 nm and the glass dielectric wave-guiding layer thickness ($d_2$) is optimized. Both the horizontal $\delta_+^H$ and vertical $\delta_+^V$ polarized PSHE with respect to the incident angle are shown in Fig. 3(a). When the Au layer is 1 nm thick and the glass cap layer is 58.2 nm, it meets the resonance condition at the incident angle of $\theta_I = 31.63°$. At this resonance condition, the horizontal polarized G-PSHE $\delta_+^H$ is $2.62 \times 10^6$ nm. The vertically polarized PSHE is $\delta_+^V$ is negligible in comparison to the horizontally polarized $\delta_+^H$ G-PSHE.

The PSHE is driven by the transverse nature of light (k.E=0), which causes the incident light to have slightly different polarization bases and a spin-dependent geometric phase. According to Eqn. (5), the transverse displacement or PSHE of reflected light for H-polarization is proportional to the ratio of $|r_s/r_p|$. The large ratio $|r_s/r_p|$ near the resonant dip will boost the H-polarization transverse shifts. To get a high $|r_s/r_p|$ ratio, the reflected light of p-polarized light, $|r_p|$, must approach a near-zero value as small as possible. Similarly, according to Eqn. (6), the transverse displacement of reflected light for V-polarization is proportional to the ratio of $|r_p/r_s|$. The significant ratio of $|r_p/r_s|$ near the resonant dip will enhance the transverse shifts of V-polarized incident light. A near-zero $|r_s|$ value will facilitate the enhanced $|r_p/r_s|$ ratio and G-PSHE.

This large G-PSHE is attributed to a high absolute Fresnel reflection coefficient ratio, $\frac{|r_s|}{|r_p|} = 1.64 \times 10^4$ (Fig. 3(b)). $|r_p|$ approaches zero in the denominator, while $|r_s|$ in the numerator is larger than $|r_p|$, that is, $|r_s|>>|r_p|$ (Fig. 3c). It is well known that the incident angle and physical parameters have significant effects on SPR generation[41]. To understand the effect

of the glass dielectric wave-guiding layer thickness ($d_2$) on the polarization and resonance angle, we performed a parameter study of $d_2$ to match the resonance condition. At the incident angle of $\theta_I = 34.81°$, another resonance condition was observed. The G-PSHE was on the order of $5 \times 10^5$ nm at a thickness of 187.6 nm of the glass dielectric wave-guiding layer with a $|r_s|/|r_p|$ ratio on the order of $3.4 \times 10^3$. Further experiments were performed by increasing the thickness of the glass dielectric wave-guiding layer. More resonance conditions appeared at glass dielectric wave-guiding layer thicknesses of 303.5 nm and 441.9 nm when the incident resonance angles were 31.63° and 34.81°, respectively. One interesting observation is that the resonance angles are alternatively periodic, that is, the third incident resonance angle is the same as the first resonance angle at 31.63° and the fourth incident resonance angle is the same as the second resonance angle at 34.81°, as shown in Table I.

**Table I.** We calculated the conditions for G-PSHE by varying the thickness of a wave-guiding layer when the Au layer thickness was fixed at 1 nm. We obtained the optimized thicknesses of the wave-guiding glass layer, the incident angles, and the polarization modes for the resonances.

| Optimized wave-guiding layer thickness ($d_2$ (nm)) | Resonance Angle ($\theta_i°$) | Resonance Polarized Mode |
|---|---|---|
| 58.2 | 31.63° | Horizontal |
| 187.6 | 34.81° | Horizontal |
| 303.5 | 31.63° | Horizontal |
| 441.9 | 34.81° | Horizontal |

For $d_1=1$ nm, only the horizontal polarization active modes were observed. In the resonance condition, one of the propagation constants of the guided wave coincided with $k_x$, and the incident wave was coupled with the guided wave in the wave-guiding layer. The reflected field intensity decreased dramatically owing to the energy transfer to the wave-guiding glass layer through the Au layer[48, 49].

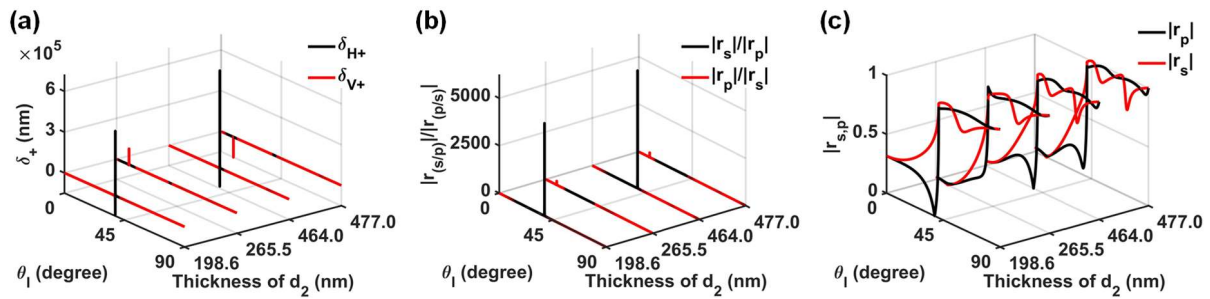

**Fig. 4.** (a) When the Au layer thickness ($d_1$) was 5 nm, by changing the thickness ($d_2$) of the wave-guiding glass layer (to 198.6 nm, 265.5 nm, 464.0 nm, and 477.0 nm), we calculated the photonic spin Hall effect of horizontal polarization $\delta_+^H$ and vertical polarization $\delta_+^V$ versus the incidence angle. (b) The absolute Fresnel reflection coefficient ratios $|r_s|/|r_p|$ and $|r_p|/|r_s|$. (c) The absolute Fresnel reflection coefficients $|r_s|$ and $|r_p|$ for the incident $|s\rangle$ polarized wave and $|p\rangle$ polarized wave with respect to the incident angle.

To understand the influence of the gold layer thickness, we also calculated the PSHE for various thicknesses of the Au layer. For a fixed 5 nm-thick ($d_1$) Au layer with a varying glass layer ($d_2$), we optimized the thickness ($d_2$) for the resonance, as shown in Fig. 4. At $d_2$=198.6 nm, SPR behavior appeared, resulting in a near-zero absolute Fresnel reflection coefficient $|r_p|$ at an incidence angle of 38.10°. The absolute Fresnel reflection coefficient $|r_s|$ was larger than $|r_p|$ at the resonance angle (Fig. 4(c)). This induced a high absolute Fresnel coefficient ratio $|r_s|/|r_p|$ (Fig. 4(b)). As a result, a horizontally polarized G-PSHE $\delta_+^H$ appeared in the order of $4.8 \times 10^5$ nm.

To probe further, we increased the thickness of the glass dielectric wave-guiding layer while fixing the 5 nm Au layer. When glass wave-guiding thickness ($d_2$) was 265.5 nm and the incident angle was maintained at $\theta_I = 9.29°$, the Fresnel coefficient ratio $|r_p|/|r_s|$ became $\sim 2 \times 10^2$. This G-PSHE was achieved for vertical polarization of the incident light. In the surface plasmon resonance (SPR) model, the typical dominant polarization mode for PSHE has always been observed for horizontal polarization because plasmonic modes only originate from TM modes[41]. However, the additional glass dielectric layer of appropriate thickness below the gold layer enables us to obtain the PSHE for both TE and TM modes owing to the guided optical wave[50].

Only p-polarized incident waves with a resonance angle are used in the popular Kretschmann configuration for local field enhancement, i.e. SPR. Previous study suggests that s-polarized field enhancement is also possible in low and high index dielectric interface[51]. The final layer must be low index layer, air in our case to maximize the intensity of evanescent wave. We proposed WG-SPR structure by combining the both waveguiding and SPR effects. This Kretschmann configuration modification allow us to enhance the evanescent wave in both s- and p- plane polarized incident waves. The enhancement of the s-polarized evanescent wave is preserved by the dielectric layer, while the p-polarized evanescent wave is enhanced by the thin metal layer in a huge scale[49, 52]. The formation of wave-guiding mode plays the key role here.

The dielectric layer deposited onto the Au thin layer acts as a waveguide layer, i.e. providing the hybrid resonance mode including plasmonic transverse magnetic mode and conventional transverse electric mode. In addition, it also significantly increases the evanescent field of both s- and p-polarized waves[49, 51, 52]. It's worth noting that the excitation of evanescent waves is quite sensitive to layer thickness or refractive index. The collective response of the described WG-SPR structure is affected by a small variation in the thickness of the dielectric layer. The change in the effective collective response of WG-SPR structure provides the different polarization mode excitation. This also provides the control over the manipulation of dominant polarization mode. The optimal thickness of the dielectric layer is decided to allow manipulation of the dominant resonance mode.

The presence of an additional dielectric glass layer allows the appearance of resonance in the TE-polarized light. Owing to the boundary condition, two-mode transverse magnetic and transverse electric modes can exist in the wave-guiding layer. An additional glass layer enhances the evanescent field near the interface between the separated media. This results in hybrid transverse magnetic and conventional waveguide transverse electric modes. By using this wave-guiding surface plasmon resonance model, we demonstrate the ability to enhance and control the evanescent field for both TE and TM modes.

From Table II, which describes a 5 nm-thick Au layer ($d_1$), a very interesting periodicity can be observed. The active polarization mode and resonance incident angle were alternatively periodic. At the wave-guiding layer thickness ($d_2$) of 198.6 nm and 464.0 nm, the horizontal polarization mode was dominantly resonant as usual at the incidence angle of 38.10°. However, when the glass dielectric wave-guiding layer thickness ($d_2$) was 265.5 nm or 477.0 nm, the resonances were achieved by the vertical polarization mode at the incidence angle of 9.29°.

**Table II.** We calculated the conditions for G-PSHE by varying the thickness of a wave-guiding layer when the Au layer thickness was fixed at 5 nm. We obtained the optimized thicknesses of the wave-guiding layer, the incident angles, and the polarization modes for the resonances.

| Optimized wave-guiding layer thickness ($d_2$ ($nm$)) | Resonance Angle ($\theta_i^\circ$) | Resonance Polarized Mode |
|---|---|---|
| 198.6 | 38.10° | Horizontal |
| 265.5 | 9.29° | Horizontal |
| 464.0 | 38.10° | Horizontal |
| 477.0 | 9.29° | Horizontal |

Using the WG-SPR model, we investigated the G-PSHE for various Au thicknesses ($d_1$) of 10 nm, 15 nm, and 25 nm versus incidence angle by varying the glass wave-guiding layer thickness ($d_2$), as presented in Fig. 5. The absolute Fresnel reflection coefficient and its corresponding ratio with other details are provided in the supporting information (Figs. S1-S5 and Table S1). A structure tolerance in terms of wave-guiding layer thickness has been calculated for each resonance condition. (Supplementry material S4).

From our obtained G-PSHE results, an interesting trend of alternating changes in the active polarization mode and incident angle was observed. The largest transverse shift has been observed for 1 nm thickness of Au layer. As the Au layer thickness is increasing the G-PSHE is relatively decreasing. It is because of the well-known fact that, the ultrathin Au layer of few nanometers possesses the extraordinary properties to provide the strong surface plasmon resonance as it acts as two-dimensional surface plasmon[53-56]. By shrinking the material size to the nanometer scale, one can significantly alter the physics of photonic and plasmonic systems[57]. The strong surface plasmon resonance is primarily caused by the surface carrier density $n_s$. In general, the bulk carrier density $n_B$ is much greater than the surface carrier density ($n_B \gg n_s$). The relationship between the two is proportional to the thickness of the metal layer, $n_s = n_B t$, (t = thickness of metal layer)[53]. To obtain $n_s$ for strong plasmon resonance, the metal layer thickness must be reduced by a few nanometers, as in graphene[58]. The ultrathin Au layer of few nanometers achieves the two-dimensional plasmon.

Ultrathin Au metal films have been discovered to be a viable solution for achieving high electro-optical tenability[53-55]. Because electrons in thin films are scarce, they are more sensitive to their surroundings, allowing electrical gating to produce large plasmon shifts in ultrathin metal layers[54]. Nowadays, with modern technology and nanostructure manufacturing, plasmonic response is available on demand[59]. Similar to graphene, Au films with few nanometer thicknesses can drastically reduce the surface plasmon wavelength $\lambda_p$[60]. Regardless of the material and thickness of the metal, the confined electric field extends a distance $\sim \lambda_p/4\pi$ away from the film in the vertical direction, whereas reduced vertical size of Au allows them to interact better with neighboring materials (dielectric layer in this study). This suggests a great opportunity for strong SPR to reduce the reflection coefficient up to near zero.

Finally, for a specific Au layer, we can manipulate the active polarization mode for giant PSHE by adding a glass wave-guiding layer. The incidence angle for the resonance can also be tuned with proper choices of physical dimensions of the presented WG-SPR model.

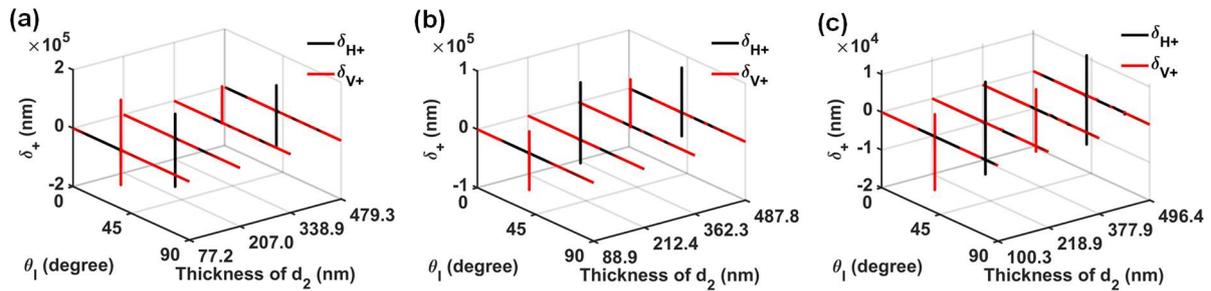

**Fig. 5.** For the Au layer with varying thicknesses of 10 nm, 15 nm, and 25 nm, by changing the thickness ($d_2$) of a wave-guiding glass layer, we calculated the optimized photonic spin Hall effect of horizontal polarization $\delta_+^H$ and vertical polarization $\delta_+^V$ versus the incidence angle. The optimized geometries are (a) $d_1$=10 nm and $d_2$=77.2 nm, 207.0 nm, and 338.9 nm, 479.3 nm. (b) $d_1$=15 nm, and $d_2$= 88.9 nm, 212.4 nm, 362.3 nm, and 487.8 nm. (c) $d_1$=25 nm, $d_2$=100.3 nm, 218.9 nm, 377.9 nm, and 496.4 nm.

To clarify the resonance caused by the vertical polarization mode, we also performed simulations for the two-dimensional equivalent structure of the presented WG-SPR model by using the wave optics module of the COMSOL Multiphysics commercial software package. As shown in Fig. 6, we calculated the field profiles of the y-component of the incident TM and TE waves to understand the role of the wave-guiding dielectric layer.

Fig. 6(a) shows the SPR field profile for the geometry when $d_1$= 25 nm and $d_2$=0, that is, a thin gold layer without a wave-guiding layer. The incident TM wave generates SPR at a resonance angle of 45.37°. Strong fields are confined at the metal-air interface. Figs. 6(b) and 6(c) also present the field profiles for SPR generation for TM waves at an incident angle of 40.20° for a thin gold layer ($d_1$= 25 nm) with a wave-guiding layer of thickness $d_2$=218.9 nm and 496.4 nm, respectively. There are strong field confinements at two interfaces, one at the metal-wave guiding layer and the other at the wave guiding layer-air interface.

Although the TE wave does not generate SPR, the presence of an additional wave guiding layer allows the feasibility of realizing a hybrid mode, that is, TM and TE mode field propagation. Figs. 6(d) and 6(e) show the field profiles, showing the strong confinement of the y-component of the TE wave at an incident angle of 41.20°, when the wave guiding layer thickness ($d_2$) is 100.3 nm and 377.9 nm, respectively. By adding a glass layer, we enhanced and controlled the evanescent field for both the TE and TM modes to achieve giant PSHEs using the WG-SPR effect.

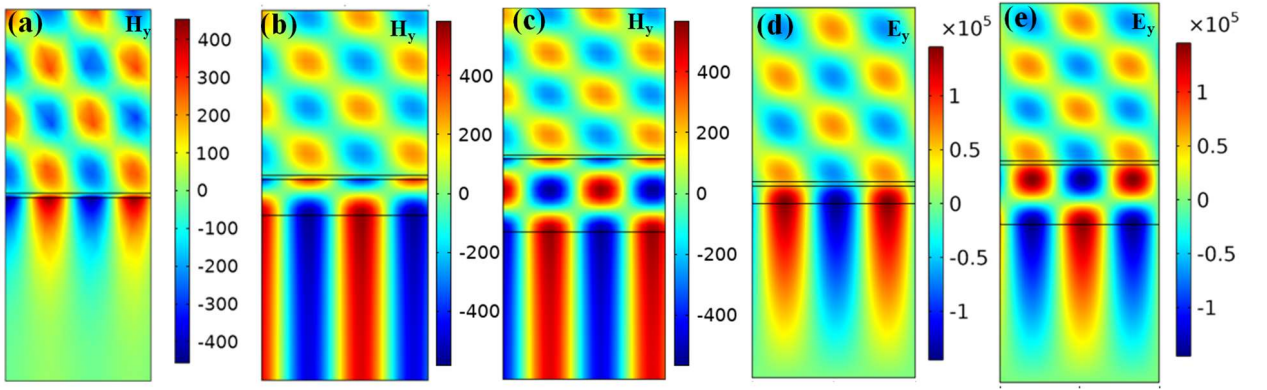

**Fig. 6.** For a fixed 25 nm-thick Au layer, we calculated the field profiles (y-component) for the incident mode (TM or TE mode) with a glass guiding layer when light was incident at the resonance angle. The geometrical parameters are (a) $d_1$=25 nm, $d_2$= 0 nm, $\theta_i = 45.37°$, TM; (b) $d_1$=25 nm, $d_2$= 218.9 nm, $\theta_i = 40.2°$, TM; (c) $d_1$=25 nm, $d_2$=496.4 nm, $\theta_i = 40.20°$, TM; (d) $d_1$=25 nm, $d_2$=100.3 nm, $\theta_i = 41.20°$, TE; (e) $d_1$=25 nm, $d_2$= 377.9 nm, $\theta_i = 41.20°$, TE.

## 4. Conclusion

In conclusion, we observed the first enhanced G-PSHE for vertically polarized waves along with horizontally polarized waves using a WG-SPR structure. By covering the metal in the Kretschmann configuration with an additional glass dielectric layer, we have achieved near zero $|r_s|$ or $|r_p|$, which leads to a high ratio of $|r_p|/|r_s|$ or $|r_s|/|r_p|$. This high absolute Fresnel coefficient ratio is responsible for realizing giant PSHE on the order of $10^6$ nm (a few millimeters to sub-millimeter), which is significantly larger than the results of other previously reported models. Our findings also indicate that we can manipulate the active polarization mode with alternate periodicity and control the enhancement of G-PSHE by tuning the thickness ($d_2$) of the wave-guiding glass dielectric layer only. This study enables the scope for potential applications of both horizontal and vertical polarized-based quantum optical devices and sensors, where light spin plays a pivotal role.

# APPENDIX A: PSHE OF REFLECTED WAVE SPECTRUM FROM WG-SPR MODEL

The WG-SPR model presented here is a two-monolayer structure coupled with a prism with refractive index $n_0$ on an air substrate. The first layer is gold (Au) with a thickness of $d_1$ and refractive index of $n_1$. The second layer is a glass dielectric wave-guiding layer with a thickness of $d_2$ and a $n_2$ refractive index that is applied to gold. The refractive index of an infinitely thick air substrate is $n_3$.

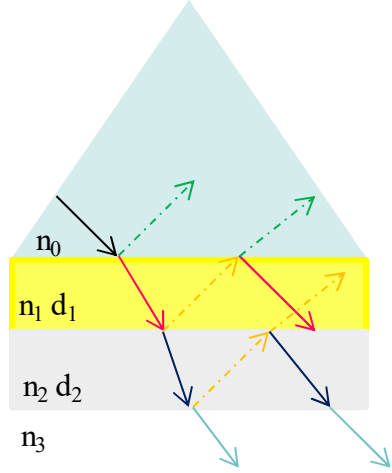

**Fig. 7.** Schematic diagram incident wave's zigzag path in wave-guiding surface Plasmon resonance structure.

By using the fundamentals of Maxwell's equations and boundary condition at interface we can develop the formulism. Due to the boundary condition, both transverse electric (TE) and transverse magnetic (TM) mode exist in waveguide. Let us first discourse for *p*-polarized (TM polarized) incident light, considering multilayer structure as shown in figure 2, thickness of gold and dielectric layer are $d_1$, $and$ $d_2$ refractive index are $n_1 and\ n_2$ corresponding to each layer, and $\theta_1, \theta_2$ are angle of incident at each interface. According to Snell's law, light propagation angle at each region is given as

$$n_0 \sin \theta_0 = n_1 \sin \theta_1 = n_2 \sin \theta_2 = n_3 \sin \theta_3 \tag{A1}$$

where numbers in subscript representing the layers: 0 in subscript is representing the initial incident medium prism at top, similarly 3 is final medium air after gold and dielectric layer.

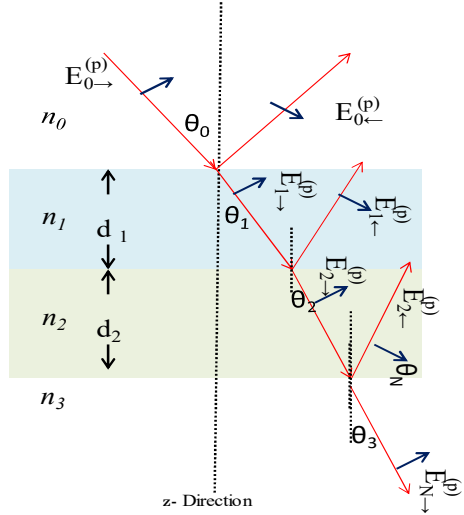

**Fig. 8.** Systematic diagram of electromagnetic wave propagation in WG-SPR model.

Acording to boundary condition we can write the field as

$$\cos\theta_0 \left(E_{0\rightarrow}^{(p)} + E_{0\leftarrow}^{(p)}\right) = \cos\theta_1 \left(E_{1\rightarrow}^{(p)} + E_{1\leftarrow}^{(p)}\right) \tag{A2}$$

and

$$n_0 \left(E_{0\rightarrow}^{(p)} - E_{0\leftarrow}^{(p)}\right) = n_0 \left(E_{1\rightarrow}^{(p)} - E_{1\leftarrow}^{(p)}\right) \tag{A3}$$

The above couple of equation is for only first interface between prism and gold layer, we can write a general equation for $j^{th}$ layer interface as,

$$\cos\theta_j \left(E_{j\rightarrow}^{(p)} e^{ik_j l_j \cos\theta_j} + E_{j\leftarrow}^{(p)} e^{-ik_j l_j \cos\theta_j}\right) = \cos\theta_{j+1} \left(E_{j+1\rightarrow}^{(p)} + E_{j+1\leftarrow}^{(p)}\right) \tag{A4}$$

and

$$n_j \left(E_{j\rightarrow}^{(p)} e^{ik_j l_j \cos\theta_j} - E_{j\leftarrow}^{(p)} e^{-ik_j l_j \cos\theta_j}\right) = n_{j+1}(E_{j+1\rightarrow}^{(p)} - E_{j+1\leftarrow}^{(p)}) \tag{A5}$$

There is no any backward field after last (dielectric) layer so the boundary condition equation at last interface we may write it as

$$\cos\theta_2 \left(E_{2\rightarrow}^{(p)} e^{ik_2 d_2 \cos\theta_2} + E_{2\leftarrow}^{(p)} e^{-ik_2 d_2 \cos\theta_2}\right) = \cos\theta_3 E_{3\rightarrow}^{(p)} \tag{A6}$$

and

$$n_2 \left(E_{2\rightarrow}^{(p)} e^{ik_2 d_2 \cos\theta_2} - E_{2\leftarrow}^{(p)} e^{-ik_2 l_2 \cos\theta_2}\right) = n_3 E_{3\rightarrow}^{(p)} \tag{A7}$$

We may summarize the above equations from 2 to 7 in a general matrix equation, given by

$$\begin{bmatrix} \cos\theta_j e^{i\beta_j} & \cos\theta_j e^{-i\beta_j} \\ n_j e^{i\beta_j} & -n_j e^{-i\beta_j} \end{bmatrix} \begin{bmatrix} E_{j\rightarrow}^{(p)} \\ E_{j\leftarrow}^{(p)} \end{bmatrix} = \begin{bmatrix} \cos\theta_{j+1} & \cos\theta_{j+1} \\ n_{j+1} & -n_{j+1} \end{bmatrix} \begin{bmatrix} E_{j+1\rightarrow}^{(p)} \\ E_{j+1\leftarrow}^{(p)} \end{bmatrix} \tag{A8}$$

Where

$$\beta_j = \begin{cases} 0, & j = 0 \\ k_j l_j \cos\theta_j, & 1 \leq j \leq N \end{cases}$$

Solution of equation 2 is given as;

$$\begin{bmatrix} E_{j\rightarrow}^{(p)} \\ E_{j\leftarrow}^{(p)} \end{bmatrix} = \begin{bmatrix} \cos\theta_j e^{i\beta_j} & \cos\theta_j e^{-i\beta_j} \\ n_j e^{i\beta_j} & -n_j e^{-i\beta_j} \end{bmatrix}^{-1} \begin{bmatrix} \cos\theta_{j+1} & \cos\theta_{j+1} \\ n_{j+1} & -n_{j+1} \end{bmatrix} \begin{bmatrix} E_{j+1\rightarrow}^{(p)} \\ E_{j+1\leftarrow}^{(p)} \end{bmatrix} \quad (A9)$$

For j=0,

$$\begin{bmatrix} E_{0\rightarrow}^{(p)} \\ E_{0\leftarrow}^{(p)} \end{bmatrix} = \begin{bmatrix} \cos\theta_0 & \cos\theta_0 \\ n_0 & -n_0 \end{bmatrix}^{-1} \begin{bmatrix} \cos\theta_1 & \cos\theta_1 \\ n_1 & -n_1 \end{bmatrix} \begin{bmatrix} E_{1\rightarrow}^{(p)} \\ E_{1\leftarrow}^{(p)} \end{bmatrix}$$

$$= \begin{bmatrix} \cos\theta_0 & \cos\theta_0 \\ n_0 & -n_0 \end{bmatrix}^{-1} M_1^{(p)} \begin{bmatrix} \cos\theta_2 & \cos\theta_2 \\ n_2 & -n_2 \end{bmatrix} \begin{bmatrix} E_{2\rightarrow}^{(p)} \\ E_{2\leftarrow}^{(p)} \end{bmatrix}$$

$$M_1^{(p)} = \begin{bmatrix} \cos\theta_0 & \cos\theta_0 \\ n_0 & -n_0 \end{bmatrix} \begin{bmatrix} \cos\theta_1 e^{i\beta_1} & \cos\theta_1 e^{-i\beta_1} \\ n_1 e^{i\beta_1} & -n_1 e^{-i\beta_1} \end{bmatrix}^{-1} \quad (A10)$$

The above equation is for j=1 combining the j=0 terms, now similarly for the j=N we can write as;

$$\begin{bmatrix} E_{0\rightarrow}^{(p)} \\ E_{0\leftarrow}^{(p)} \end{bmatrix} = \begin{bmatrix} \cos\theta_0 & \cos\theta_0 \\ n_0 & -n_0 \end{bmatrix}^{-1} \left( \prod_{j=1}^{N} M_j^{(p)} \right) \begin{bmatrix} \cos\theta_3 & \cos\theta_3 \\ n_3 & -n_3 \end{bmatrix} \begin{bmatrix} E_{3\rightarrow}^{(p)} \\ 0 \end{bmatrix} \quad (A11)$$

$$M_j^{(p)} = \begin{bmatrix} \cos\theta_j & \cos\theta_j \\ n_j & -n_j \end{bmatrix} \begin{bmatrix} \cos\theta_j e^{i\beta_j} & \cos\theta_j e^{-i\beta_j} \\ n_j e^{i\beta_j} & -n_j e^{-i\beta_j} \end{bmatrix}^{-1}$$

$$= \begin{bmatrix} \cos\beta_j & -i\sin\beta_j \cos\theta_j / n_j \\ -i n_j \sin\beta_j / \cos\theta_j & \cos\beta_j \end{bmatrix} \quad (A12)$$

By using the matrix inversion property, the solution we may write it as,

$$\begin{bmatrix} 1 \\ E_{0\leftarrow}^{(p)} / E_{0\rightarrow}^{(p)} \end{bmatrix} = A^{(p)} \begin{bmatrix} E_{3\rightarrow}^{(p)} / E_{0\rightarrow}^{(p)} \\ 0 \end{bmatrix} \quad (A13)$$

where,

$$A^{(p)} = \begin{bmatrix} a_{11}^{(p)} & a_{12}^{(p)} \\ a_{21}^{(p)} & a_{22}^{(p)} \end{bmatrix} = \frac{1}{2n_0 \cos\theta_0} \begin{bmatrix} n_0 & \cos\theta_0 \\ n_0 & -\cos\theta_0 \end{bmatrix} \left( M_1^{(p)} \times M_2^{(p)} \right) \begin{bmatrix} \cos\theta_3 & 0 \\ n_3 & 0 \end{bmatrix} \quad (A14)$$

$$r_p = \frac{E_{0\leftarrow}^{(p)}}{E_{0\rightarrow}^{(p)}} = \frac{a_{21}^{(p)}}{a_{11}^{(p)}} \quad (A15)$$

The all above stated equations are for *p-polarized* light, similarly we can write all equations for *s-polarized* light

$$\begin{bmatrix} 1 \\ E_{0\leftarrow}^{(s)} / E_{0\rightarrow}^{(s)} \end{bmatrix} = A^{(s)} \begin{bmatrix} E_{3\rightarrow}^{(s)} / E_{0\rightarrow}^{(s)} \\ 0 \end{bmatrix} \quad (A16)$$

$$A^{(s)} = \begin{bmatrix} a_{11}^{(s)} & a_{12}^{(s)} \\ a_{21}^{(s)} & a_{22}^{(s)} \end{bmatrix} = \frac{1}{2n_0 cos\theta_0} \begin{bmatrix} n_0 cos\theta_0 & 1 \\ -n_0 cos\theta_0 & -1 \end{bmatrix} \left(M_1^{(s)} \times M_2^{(s)}\right) \begin{bmatrix} 1 & 0 \\ n_3 cos\theta_3 & 0 \end{bmatrix} \quad (A17)$$

$$M_j^{(s)} = \begin{bmatrix} cos\beta_j & -isin\beta_j/n_j cos\theta_j \\ -in_j sin\beta_j cos\theta_j & cos\beta_j \end{bmatrix} \quad (A18)$$

$$r_s = \frac{E_{0\leftarrow}^{(s)}}{E_{0\rightarrow}^{(s)}} = \frac{a_{21}^{(s)}}{a_{11}^{(s)}} \quad (A19)$$

In the spin bias set representation, the incident monochromatic wave can be shown in two polarization state.

$$\widetilde{E}_i^H = \frac{1}{\sqrt{2}}\left(\widetilde{E}_{i|+>} + \widetilde{E}_{i|->}\right)$$
$$\widetilde{E}_i^V = \frac{i}{\sqrt{2}}\left(\widetilde{E}_{i|->} - \widetilde{E}_{i|+>}\right) \quad (A20)$$

Where H and V stands for horizontal and vertical polarized wave. $\widetilde{E}_{i|+>}$ and $\widetilde{E}_{i|->}$ is representing left-handed and right-handed circularly polarized wave. The spectrum of incident light is extremely narrow and given as

$$\widetilde{E}_{i|\pm>} = (\vec{e}_{ix} + i\sigma\vec{e}_{iy})\frac{\omega_0}{\sqrt{2\pi}}e^{-\left(\frac{\omega_0^2(k_{ix}^2 + k_{iy}^2)}{4}\right)} \quad (A21)$$

Where $\omega_0$ is light waist, $\sigma = \pm 1$ plan operators, $\sigma = +1$ and $\sigma = -1$ are presented left circular polarized light and right circularly polarized, respectively. Due to the matching of boundary condition, the reflected light can be represented as

$$\begin{bmatrix} \widetilde{E}_r^H \\ \widetilde{E}_r^V \end{bmatrix} = \begin{bmatrix} r_p & \frac{k_{ry}(r_p + r_s)cot\theta_i}{k_0} \\ -\frac{k_{ry}(r_p + r_s)cot\theta_i}{k_0} & r_s \end{bmatrix} \begin{bmatrix} \frac{1}{\sqrt{2}}\left(\widetilde{E}_{i|+>} + \widetilde{E}_{i|->}\right) \\ \frac{i}{\sqrt{2}}\left(\widetilde{E}_{i|+>} + \widetilde{E}_{i|->}\right) \end{bmatrix} \quad (A22)$$

$$\widetilde{E}_{r|\pm>}^H = \frac{(e_{rx} \pm ie_{ry})}{\sqrt{\pi}\omega_0} \frac{z_R}{z_R + iz_r} e^{\left[-\frac{k_0}{2}\frac{x_r^2 + y_r^2}{z_R + iz_r}\right]}$$
$$\times \left[r_p - \frac{ix}{z_R + iz_r}\frac{\partial r_p}{\partial\theta_i} \pm \frac{y}{z_R + iz_r}(r_p + r_s)\right.$$
$$\left. \pm \frac{ixy}{(z_R + iz_r)^2}\left(\frac{\partial r_p}{\partial\theta_i} + \frac{\partial r_s}{\partial\theta_i}\right)\right]e^{(ik_r z_r)}$$
$$(A23)$$

$$\widetilde{E}_{r|\pm>}^V = \frac{\mp(e_{rx} \pm ie_{ry})}{\sqrt{\pi}\omega_0} \frac{z_R}{z_R + iz_r} e^{\left[-\frac{k_0}{2}\frac{x_r^2 + y_r^2}{z_R + iz_r}\right]}$$
$$\times \left[r_s - \frac{ix}{z_R + iz_r}\frac{\partial r_s}{\partial\theta_i} \pm \frac{y}{z_R + iz_r}(r_p + r_s)\right.$$
$$\left. \pm \frac{ixy}{(z_R + iz_r)^2}\left(\frac{\partial r_p}{\partial\theta_i} + \frac{\partial r_s}{\partial\theta_i}\right)\right]e^{(ik_r z_r)}$$
$$(A24)$$

$$\widetilde{E}_r^H = \frac{r_p}{\sqrt{2}}\left[e^{(ik_{ry}\delta_{r|+>}^H)}\widetilde{E}_{r|+>} + e^{(-ik_{ry}\delta_{r|->}^H)}\widetilde{E}_{r|->}\right] \quad (A25)$$

$$\widetilde{E}_r^V = i\frac{r_s}{\sqrt{2}}\left[-e^{(ik_{ry}\delta_{r|+>}^V)}\widetilde{E}_{r|+>} + e^{(-ik_{ry}\delta_{r|->}^V)}\widetilde{E}_{r|->}\right] \qquad (A26)$$

where $z_R = k_0\omega_0^2/2$ is the Rayleigh length and $e^{(\pm ik_{ry}\delta_r^{H,V})}$, term is responsible for spin-orbit interaction of light.

Where,

$$\delta_{r|\pm>}^H = \mp\left(1 - \frac{2r_s}{r_p} + \left(\frac{r_s}{r_p}\right)^2\right)\frac{\cot\theta_i}{k_0}$$

$$\delta_{r|\pm>}^V = \mp\left(1 - \frac{2r_p}{r_s} + \left(\frac{r_p}{r_s}\right)^2\right)\frac{\cot\theta_i}{k_0}$$

$$\delta_{r|\pm>}^H = \mp\frac{\lambda_0}{2\pi}\left[1 + \frac{|r_s|}{|r_p|}\cos(\varphi_s - \varphi_p)\right]\cot\theta_i \qquad (A27)$$

$$\delta_{r|\pm>}^V = \mp\frac{\lambda_0}{2\pi}\left[1 + \frac{|r_p|}{|r_s|}\cos(\varphi_p - \varphi_s)\right]\cot\theta_i \qquad (A28)$$

In the above expression $r_{s,p}$ is Fresnel reflection coefficient for $|s>$ and $|p>$ polarized incident wave respectively and $\varphi_{s,p}$ is corresponding phase. $\delta_{r|\pm>}^H$ and $\delta_{r|\pm>}^V$ are measure of shift of left- and right-handed circular polarization component of H and V polarized light.

# Supplementary Material

# Polarization manipulation of giant photonic spin Hall effect using wave-guiding effect

Monu Nath Baitha and Kyoungsik Kim[*]

School of Mechanical Engineering, Yonsei University, 50 Yonsei-ro, Seodaemun-gu, Seoul 03722, Republic of Korea.

[*]Correspondence and requests for materials should be addressed to K.K. (email: kks@yonsei.ac.kr).

## S1. Calculated giant-photonics spin Hall Effect (G-PSHE) for different geometry of wave-guided surface plasmonic resonance (WG-SPR) model

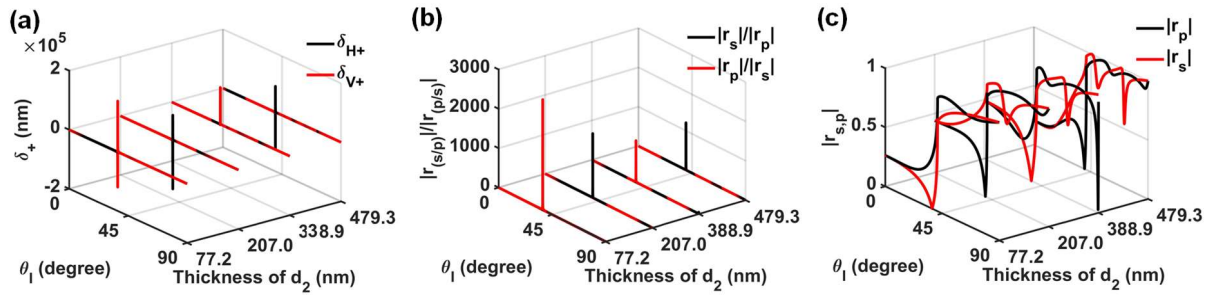

**Fig. S1.** (a) When the Au layer thickness ($d_1$) was 10 $nm$, by changing the thickness ($d_2$) of the wave-guiding glass layer (to 77.2 $nm$, 207.0 $nm$, 338.9 $nm$, and 479.3 $nm$), we calculated the photonic spin Hall effect of horizontal polarization $\delta_+^H$ and vertical polarization $\delta_+^V$ versus the incidence angle. (b) The absolute Fresnel reflection coefficient ratios $|r_s|/|r_p|$ and $|r_p|/|r_s|$. (c) The absolute Fresnel reflection coefficients $|r_s|$ and $|r_p|$ for the incident |s> polarized wave and |p> polarized wave with respect to the incident angle.

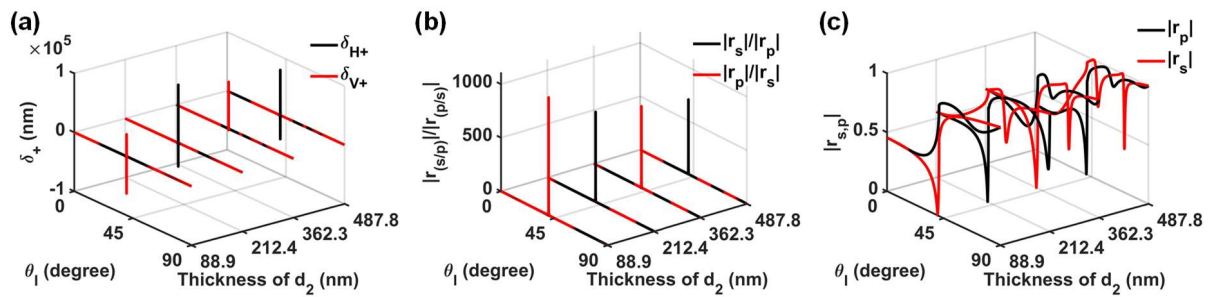

**Fig. S2.** (a) When the Au layer thickness ($d_1$) was 15 $nm$, by changing the thickness ($d_2$) of the wave-guiding glass layer (to 88.9 $nm$, 212.4 $nm$, 362.3 $nm$, and 487.8 $nm$), we calculated the photonic spin Hall effect of horizontal polarization $\delta_+^H$ and vertical polarization $\delta_+^V$ versus the incidence angle. (b) The absolute Fresnel reflection coefficient ratios $|r_s|/|r_p|$ and $|r_p|/|r_s|$. (c)

The absolute Fresnel reflection coefficients |$r_s$| and |$r_p$| for the incident |s⟩ polarized wave and |p⟩ polarized wave with respect to the incident angle.

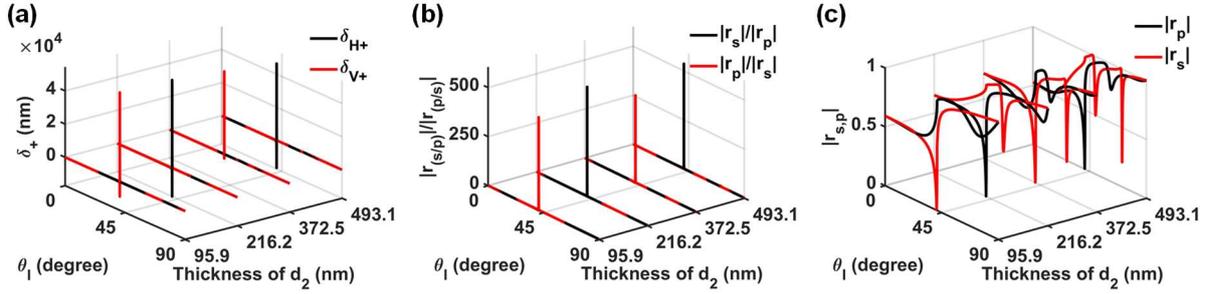

**Fig. S3.** (a) When the Au layer thickness ($d_1$) was 20 $nm$, by changing the thickness ($d_2$) of the wave-guiding glass layer (to 95.9 $nm$, 216.2 $nm$, 372.5 $nm$, and 493.1 $nm$), we calculated the photonic spin Hall effect of horizontal polarization $\delta_+^H$ and vertical polarization $\delta_+^V$ versus the incidence angle. (b) The absolute Fresnel reflection coefficient ratios |$r_s$|/|$r_p$| and |$r_p$|/|$r_s$|. (c) The absolute Fresnel reflection coefficients |$r_s$| and |$r_p$| for the incident |s⟩ polarized wave and |p⟩ polarized wave with respect to the incident angle.

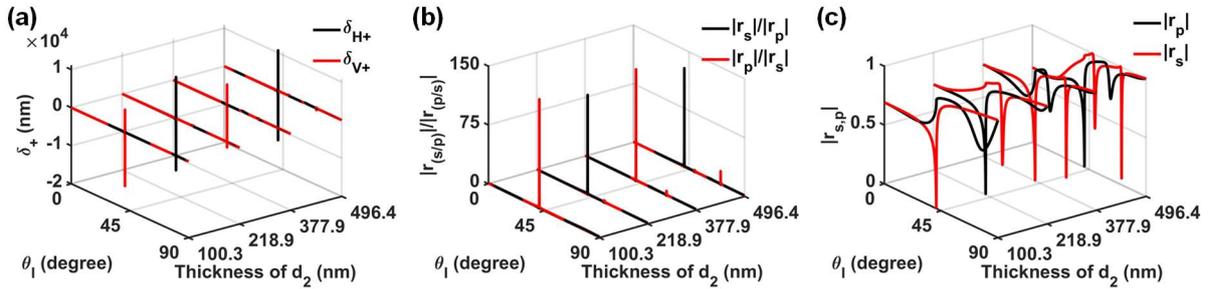

**Fig. S4.** (a) When the Au layer thickness ($d_1$) was 25 $nm$, by changing the thickness ($d_2$) of the wave-guiding glass layer (to 100.3 $nm$, 218.9 $nm$, 377.9 $nm$, and 496.4 $nm$), we calculated the photonic spin Hall effect of horizontal polarization $\delta_+^H$ and vertical polarization $\delta_+^V$ versus the incidence angle. (b) The absolute Fresnel reflection coefficient ratios |$r_s$|/|$r_p$| and |$r_p$|/|$r_s$|. (c) The absolute Fresnel reflection coefficients |$r_s$| and |$r_p$| for the incident |s⟩ polarized wave and |p⟩ polarized wave with respect to the incident angle.

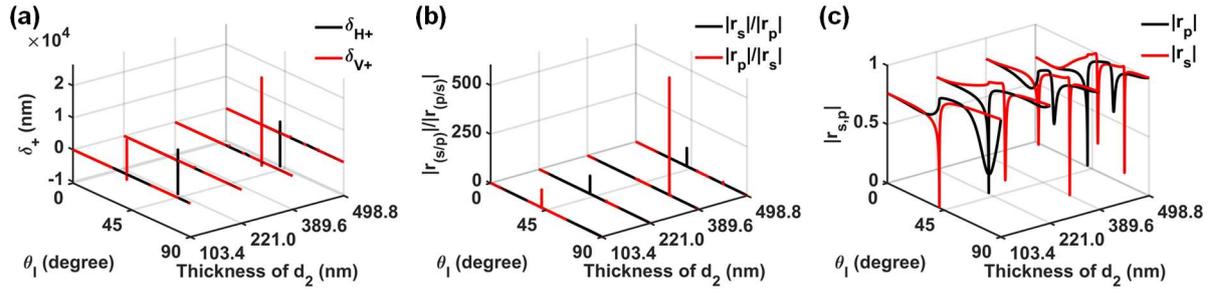

**Fig. S5.** (a) When the Au layer thickness ($d_1$) was 30 $nm$, by changing the thickness ($d_2$) of the wave-guiding glass layer (to 103.4 $nm$, 221.0 $nm$, 389.6 $nm$, and 498.8 $nm$), we calculated the photonic spin Hall effect of horizontal polarization $\delta_+^H$ and vertical polarization $\delta_+^V$ versus the incidence angle. (b) The absolute Fresnel reflection coefficient ratios $|r_s|/|r_p|$ and $|r_p|/|r_s|$. (c) The absolute Fresnel reflection coefficients $|r_s|$ and $|r_p|$ for the incident |s> polarized wave and |p> polarized wave with respect to the incident angle.

### S2. Summary of the optimized thickness of glass dielectric wave-guiding layer, resonance angle, and active polarization mode for different Au thicknesses.

Using the WG-SPR model, we investigated the giant PSHE for various Au thicknesses ($d_1$) of 1 nm to 30 $nm$ versus incidence angle by varying the glass wave-guiding layer thickness ($d_2$). For the given thickness of the gold layer, we obtained the resonance conditions, such as dielectric layer thickness, incidence angle, and active polarization modes.

**TABLE SI:** Details of optimize resonance thickness, resonance angle and active polarized mode for Au 10 $nm$, 15 $nm$, 20 $nm$, 25 $nm$, and 30 $nm$.

| Geometry | Au Thickness $d_1$ ($nm$) | Glass dielectric wave-guiding optimized thickness $d_2$ ($nm$) | Resonance incident angle $\theta_I$ (degree) | Active polarization mode |
|---|---|---|---|---|
| 1 | 10 | 77.2 | 37.05° | Vertical |
|   |    | 207.0 | 39.91° | Horizontal |
|   |    | 338.9 | 37.05° | Vertical |
|   |    | 479.3 | 39.91° | Horizontal |
| 2 | 15 | 88.9 | 40.18° | Vertical |
|   |    | 212.4 | 40.69° | Horizontal |
|   |    | 362.3 | 40.18° | Vertical |
|   |    | 487.8 | 40.69° | Horizontal |
| 3 | 20 | 95.9 | 40.97° | Vertical |
|   |    | 216.2 | 41.04° | Horizontal |
|   |    | 372.5 | 40.97° | Vertical |

| | | 493.1 | 41.04° | Horizontal |
|---|---|---|---|---|
| | | | | |
| 4 | 25 | 100.3 | 41.20° | Vertical |
| | | 218.9 | 40.20° | Horizontal |
| | | 377.9 | 41.20° | Vertical |
| | | 496.4 | 40.20° | Horizontal |
| | | | | |
| 5 | 30 | 103.4 | 41.28° | Vertical |
| | | 221.0 | 41.26° | Horizontal |
| | | 389.6 | 66.27° | Vertical |
| | | 498.8 | 41.26° | Horizontal |

## S3. COMSOL simulation study

To explore the impact of glass dielectric wave-guiding layer on G-PSHE, the simulations were carried out based on finite element method (FEM) by using COMSOL commercial software package. Simulation has been done on equivalent 2-D structure of presented WG-SPR structure. The excitation source is a |p> and |s> polarized plane wave of 632.8 nm wavelength at an incident angle from top of prim. In addition, the Floquet periodic boundary conditions (along the $+x$ and $-x$-axis directions) were applied for the infinite extension of structure. The $E_y$ and $H_y$ filed profile has been calculated for TE and TM polarization mode respectively.

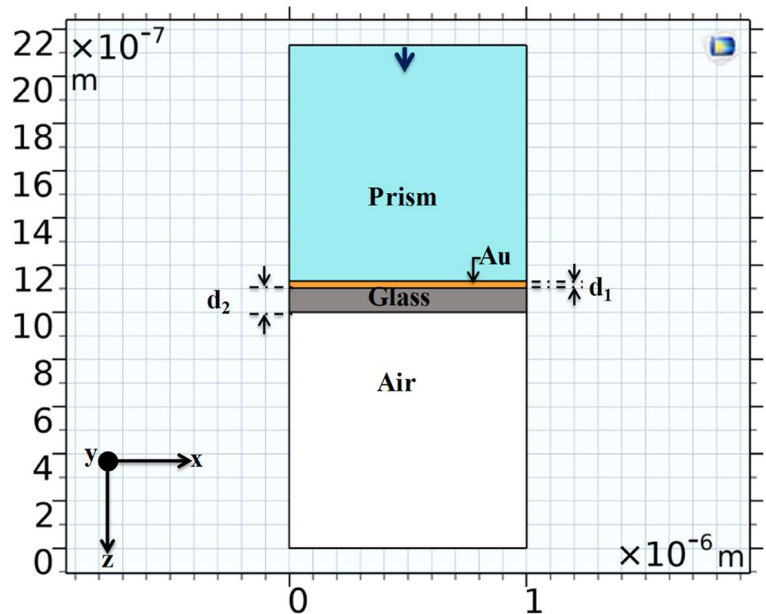

FIG. S6. 2-D structure of presented WG-SPR structure for simulation

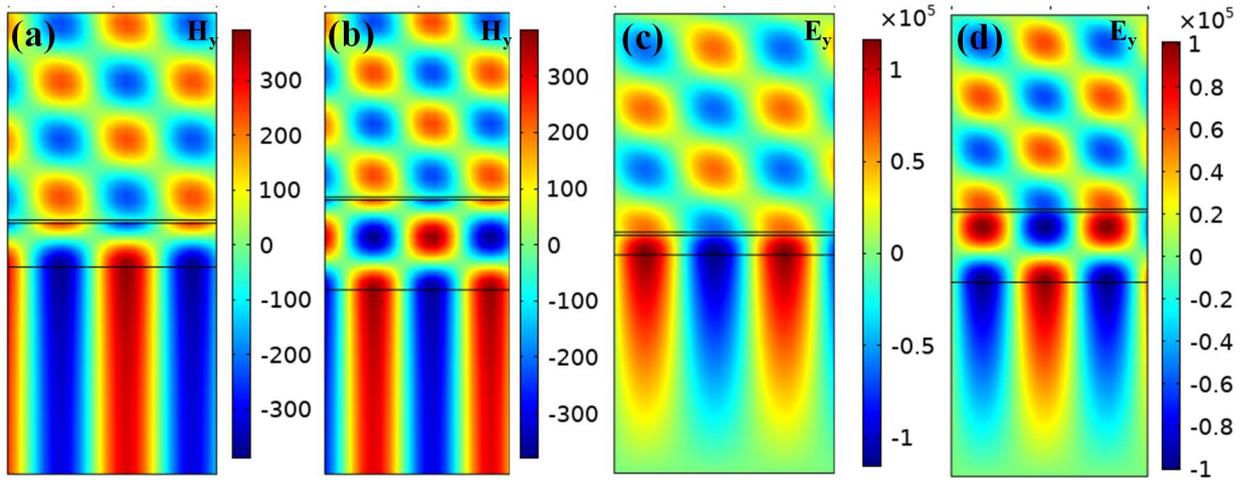

**Fig. S7**. For a fixed 15 $nm$-thick Au layer, we calculated the field profiles (y-component) for the incident mode (TM or TE) with a glass guiding layer when light is incident at resonance angle. The geometrical parameters are (a) $d_1$=15 $nm$, $d_2$= 212.4 $nm$, $\theta_i = 40.69°$, TM (b) $d_1$=15 $nm$, $d_2$= 487.8 $nm$, $\theta_i = 40.69°$, TM (c) $d_1$=15 $nm$, $d_2$= 88.9 $nm$, $\theta_i = 40.18°$, TE (d) $d_1$=15 $nm$, $d_2$=362.3 $nm$, $\theta_i = 40.18°$, TE.

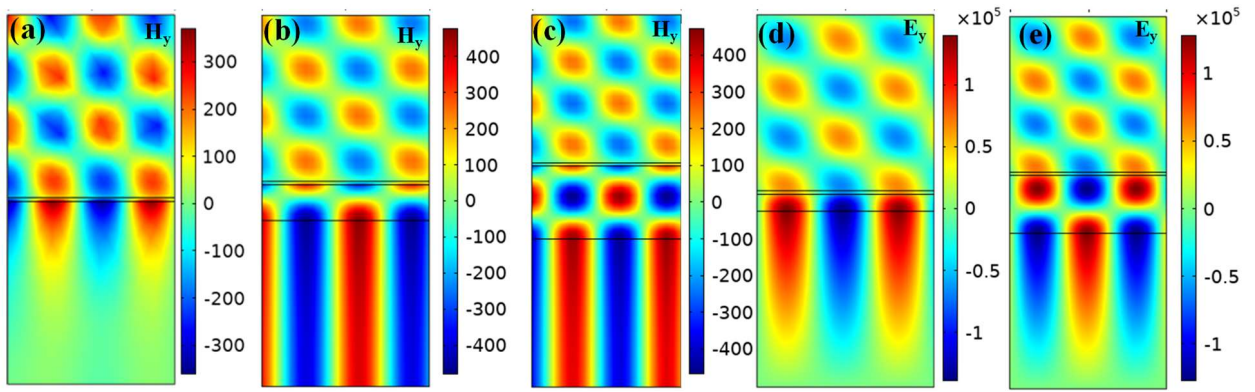

**Fig. S8**. For a fixed 20 $nm$-thick Au layer, we calculated the field profiles (y-component) for the incident mode (TM or TE) with a glass guiding layer when light is incident at resonance angle. The geometrical parameters are The geometrical parameters are (a) $d_1$=20 $nm$, $d_2$= 0 $nm$, $\theta_i = 45.09°$, TM (b) $d_1$=20 $nm$, $d_2$= 216.2 $nm$, $\theta_i = 41.04°$, TM (c) $d_1$=20 $nm$, $d_2$= 493.1 $nm$, $\theta_i = 41.04°$, TM (d) $d_1$=20 $nm$, $d_2$=95.9 $nm$, $\theta_i = 40.97°$, TE (e) $d_1$=20 $nm$, $d_2$=372.5 $nm$, $\theta_i = 40.97°$, TE.

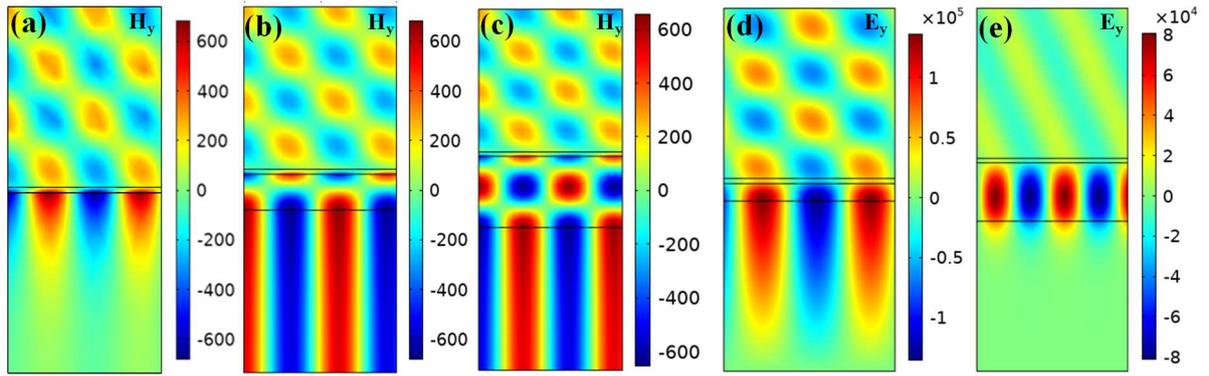

**Fig. S9.** For a fixed 30 $nm$-thick Au layer, we calculated the field profiles (y-component) for TM or TE mode with varying glass guiding layer when light is incident at resonance angle. The geometrical parameters are (a) $d_1$=30 $nm$, $d_2$= 0 $nm$, $\theta_i = 44.60°$, TM (b) $d_1$=30 $nm$, $d_2$= 221.0 $nm$, $\theta_i = 41.26°$, TM (c) $d_1$=30 $nm$, $d_2$= 498.8 $nm$, $\theta_i = 41.28°$, TM (d) $d_1$=30 $nm$, $d_2$=103.4 $nm$, $\theta_i = 41.28°$, TE (e) $d_1$=30 $nm$, $d_2$=389.6 $nm$, $\theta_i = 66.27°$, TE.